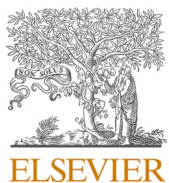
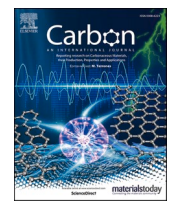

# Trion emission from frozen p-n junctions in networks of electrolyte-gated (6,5) single-walled carbon nanotubes

Abdurrahman Ali El Yumin, Nicolas F. Zorn, Felix J. Berger, Daniel Heimfarth, Jana Zaumseil[*]

*Institute for Physical Chemistry, Universität Heidelberg, 69120 Heidelberg, Germany*

## ABSTRACT

We demonstrate exciton and charged exciton (trion) electroluminescence from frozen p-n junctions in networks of polymer-sorted, semiconducting (6,5) single-walled carbon nanotubes. Electrolyte-gating with an ionic liquid was employed to achieve injection and accumulation of high densities of holes and electrons in the nanotube network at low applied voltages. Static p-n junctions were formed by cooling the devices below the melting point of the ionic liquid while in the ambipolar regime. These frozen junctions showed diode-like rectification and enabled the investigation of electron-hole recombination and near-infrared electroluminescence under controlled conditions. The contributions of exciton and red-shifted trion emission to the electroluminescence spectra were influenced by the initial parameters of the p-n junction formation (balanced or unbalanced) and the applied lateral bias, but did not depend on temperature (30–200 K). The tilted potential profile along the fixed junction and consequently the number of excess carriers within the recombination zone were found to predominantly determine the emission intensity and observed trion to exciton ratio.

## 1. Introduction

Trions are negatively or positively charged excitons. They are considered to be quasi-particles consisting of two electrons and one hole or two holes and one electron [1]. Trions can be created by optical excitation of electrostatically, electrochemically or chemically doped low-dimensional semiconductors such as quantum wells or (colloidal) quantum dots [2,3], monolayered transition metal dichalcogenides (TMDs) [4–7], other 2D semiconductors (*e.g.*, black phosphorous or ReS$_2$) [8,9], CdS nanoplatelets [10,11], and semiconducting single-walled carbon nanotubes (SWCNTs) [12–14]. They are typically observed as red-shifted peaks or shoulders in the corresponding photoluminescence (PL) spectra. The characteristic energy offset of trion emission compared to the excitonic transition is directly associated with their binding energy. Both exciton and trion binding energies increase as the semiconductor's dimensionality and thus its dielectric screening is reduced [15]. Hence, trions in 2D semiconductors typically exhibit small binding energies of 10–40 meV [16], while one-dimensional SWCNTs show much larger and diameter-dependent trion binding energies of 100–200 meV [13,14]. This large energy difference of SWCNT trions with respect to their corresponding excitons and the narrow linewidths of both enable their clear and unambiguous assignment even at room temperature [12–14,17,18]. They are easily observed for intentionally or unintentionally doped individual nanotubes [19] but also for ensembles of nanotubes with one or few different diameters and chiralities [12,20,21]. For example, (6,5) SWCNTs with a diameter of 0.76 nm exhibit trion emission at ~1180 nm while excitonic emission occurs at ~1020 nm, indicating a large energy offset of ~165 meV [17,21]. Trion emission can further serve as an optical indicator of the doping state of a nanotube sample [19,22–24]. As the concentration of charges in nanotubes and nanotube networks can be varied by electrostatic or electrochemical gating, trion emission can also be tuned, which has enabled the demonstration of trion-polaritons in optical cavities, *i.e.*, strong light-matter coupling [25].

More often however, trions are an unwanted side effect in SWCNT electroluminescent devices (*e.g.*, light-emitting transistors and diodes), leading to additional emission peaks [17,26–30]. As described above, they are indicative of excess charge carriers that lead to Auger quenching [31] and thus reduce the overall emission efficiency of these devices. So far, there are no clear guidelines or studies regarding enhancement or suppression of trions in optoelectronic devices. Further investigations regarding their dependence on absolute carrier density, dielectric environment and temperature [28] are needed. A versatile tool for such studies is the light-emitting field-effect transistor (LEFET).

SWCNTs are particularly well-suited as materials for LEFETs in general due to their small bandgap, ambipolar character, and large hole and electron mobilities [32]. LEFETs were demonstrated for networks of semiconducting nanotubes [33–38] as well as individual SWCNTs [39–41]. In an ambipolar LEFET the drain ($V_D$) and gate ($V_G$) voltages are chosen such that electrons are injected from the source electrode and holes are injected from the drain electrode (or *vice versa*) into the channel and form accumulation layers that meet at some point along the channel, where the charges recombine and light emission takes place (recombination and emission zone) [42]. By changing $V_D$ or $V_G$, the recombination zone can be moved through the entire channel. The hole and electron densities along the length of the channel depend on the local potential and ideally should reach zero in the recombination and emission zone. However, several studies - including in-situ Raman mapping of the 2D Raman mode position - have shown that this is not the case for SWCNT LEFETs and quenching of excitons by excess charges and trion emission occur [17,35].

While trion emission is barely observed in purely electrostatically gated LEFETs [43], it becomes rather dominant in electrolyte-gated LEFETs [17], including those based on transition metal dichalcogenides [44]. In an electrolyte-gated transistor the dielectric layer is replaced with an electrolyte such as an ionic liquid or an iongel with mobile cations and anions. The application of a gate voltage leads to movement of these ions in the electric field and formation of electric double layers around the carbon nanotubes and consequently accumulation of charge carriers with the opposite polarity to the respective ions [45]. Due to the nm-thick electric double layer of the ions, the effective capacitance (several $\mu F \cdot cm^{-2}$) and induced charge carrier density ($\sim 10^{14}$ $cm^{-2}$) are extremely high [46], which also explains the observation of trions.

Importantly, electrolyte-gating with ionic liquids (ILs, *e.g.*, various imidazolium salts) provides the unique opportunity to immobilize/freeze the ions by cooling the device below the melting point or glass transition temperature of the IL under an applied bias. The ion distribution and hence doping profile of the channel become fixed and independent of the applied voltages. The creation of a static junction enables a more detailed investigation of the impact of charge carrier balance and applied electric fields on trion electroluminescence. Indeed, electroluminescence (EL) from frozen p-n junctions was previously observed for TMDs [47,48] and revealed circularly polarized trion emission from WS$_2$ and MoSe$_2$ layers depending on the applied lateral bias across the junction [44,49].

Here we report trion and exciton electroluminescence from static p-n junctions in networks of (6,5) SWCNTs that were created by freezing ionic liquids under applied bias. The diode-like characteristics of the p-n junctions enable the investigation of EL, PL and the recombination zone under controlled conditions, low and variable temperature (<200 K) and additional lateral electric fields. The ratio of trion to exciton emission depends strongly on the initial balance of charge carriers around the recombination zone as well as the applied external bias, but not on temperature.

## 2. Experimental methods

### 2.1. Preparation of (6,5) SWCNT dispersions

The (6,5) SWCNT dispersions in toluene were prepared from CoMoCAT raw material (CHASM Advanced Materials Inc., SG65i-L58, 0.4 g L$^{-1}$) *via* shear force mixing (Silverson L2/Air mixer, 10230 rpm, 72 h) and polymer-wrapping with poly[(9,9-dioctylfluorenyl-2,7-diyl)-*alt*-(6,6'-(2,2'-bipyridine))] (PFO-BPy, American Dye Source, $M_W = 40$ kg mol$^{-1}$, 0.5 g L$^{-1}$) as described previously [50]. The nanotube dispersions were centrifuged twice for 45 min at 60000*g* (Beckman Coulter Avanti J26XP centrifuge) and subsequently filtered through a polytetrafluoroethylene (PTFE) syringe filter (pore size 5 μm) to remove aggregates. After that, they were filtered through a PTFE membrane (Merck Millipore JVWP; pore size 0.1 μm) to collect the nanotubes and the obtained filter cake was washed with hot toluene (80 °C) to remove the excess polymer. Finally, the filter cakes of (6,5) SWCNTs were redispersed in fresh toluene by bath sonication for 30 min.

### 2.2. Device fabrication

The electrodes (side-gate electrode pad and interdigitated source-drain electrodes with channel width $W = 10$ mm and channel length $L = 20$ μm) were patterned by photolithography (double-layer LOR5B/S1813 resist, microresist technology GmbH) on glass substrates (Schott AG, AF32eco, 300 μm thickness). Then, 2 nm of chromium and 40 nm of gold were deposited *via* electron beam evaporation followed by lift-off in *N*-methyl pyrrolidone (NMP). The substrates were cleaned by ultra-sonication in acetone and 2-propanol for 10 min each. The prepared (6,5) SWCNT dispersions with an optical density of 3.6 cm$^{-1}$ at 996 nm (E$_{11}$ transition) were spin-coated (3 × 70 μL, 2000 rpm, 30 s) onto the substrates with annealing steps (100 °C, 2 min) in between. For a representative absorption spectrum of the nanotube dispersion and atomic force micrograph of deposited networks see Fig. S1 (Supplementary Material). To remove residual polymer, the spin-coated substrates were rinsed with tetrahydrofuran and 2-propanol for 10 s each. The nanotube networks were patterned using photolithography (see above) and oxygen plasma etching to remove all SWCNTs outside the channel area. Finally, the samples were annealed at 150 °C for 30 min in a dry nitrogen atmosphere, and the ionic liquid 1-ethyl-3-methyl-imidazolium-tris(pentafluoroethyl)-trifluorophosphate ([EMIM][FAP] from Merck Millipore) was drop-cast onto the channel area and the side gate as the electrolyte. A thin glass slide was placed on top to evenly distribute the ionic liquid. This prevents cracking of the ionic liquid during cooling and improves the optical image quality.

### 2.3. Electroluminescence and photoluminescence spectroscopy

All experiments were carried out in a closed-loop liquid helium optical cryostat (Montana Instruments Cryostation s50) under high-vacuum conditions ($\sim 10^{-6}$ - $10^{-5}$ mbar). Electrical transport measurements were performed with a semiconductor parametric analyzer (Agilent 4155C). Samples were cooled below the glass transition temperature of the ionic liquid ($T \leq 240$ K, cooling rate $\sim 3$ K per minute) while specific drain and gate voltages were maintained to create frozen p-n junctions. Temperatures during EL and PL measurements were kept constant and were varied between 30 K and 200 K as noted for the corresponding experiments.

For PL measurements, samples were excited with a continuous-wave laser diode (OBIS Coherent, 640 nm, 0.5–10 mW output power). Optical imaging and collection of EL and PL signals were enabled by an infrared-optimized ×50 long working distance objective (Mitutoyo, LCD Plan Apo NIR, numerical aperture NA = 0.42) mounted outside the cryostat. A thermoelectrically cooled InGaAs camera (NIRvana 640ST) coupled to an IsoPlane SCT-320 spectrograph (Princeton Instruments) was used to acquire all emission images and spectra. For PL quenching experiments, an additional bi-concave lens was used to expand the excitation laser beam.

## 3. Results and discussion

### 3.1. Formation of p-n junction and diode-like characteristics

To investigate the EL from a static p-n junction in a nanotube network, ionic liquid-gated light-emitting field-effect transistors (IL-LEFETs) with a spin-coated and patterned layer of polymer-sorted (6,5) SWCNTs were fabricated as described above. Fig. 1a shows the schematic layout of such a device and measurement setup. The transfer characteristics of a representative IL-LEFET at room temperature (Fig. 1b) show clear ambipolar charge transport. Hole and electron accumulation are evident for negative and positive gate voltages ($V_G$), respectively, at low drain voltages ($V_D = -0.5$ V). The (6,5) SWCNTs were slightly p-doped as evident from the shift of the onset voltage for hole transport to positive $V_G$. Nevertheless, the transconductances for negative and positive gate voltages were very similar and the corresponding hole and electron mobilities were balanced as expected for semiconducting carbon nanotubes [38].

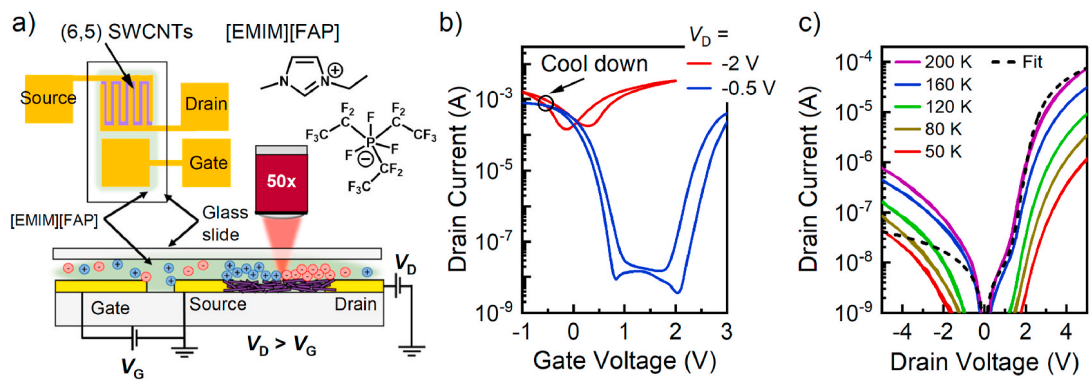

**Fig. 1.** **a)** Schematic layout and operation of an ionic liquid-gated LEFET with [EMIM][FAP] as ionic liquid and (6,5) SWCNTs. **b)** Transfer curves of IL-LEFET at low (unipolar regime, blue) and high (ambipolar regime, red) drain voltages. To form a static p-n junction, IL-LEFETs were cooled down with applied voltages (*e.g.*, $V_D = -2$ V and $V_G = -0.3$ V, black circle). **c)** I–V characteristics of frozen p-n junction (diode-like) at different temperatures below the melting point of the IL. The dashed line represents a fit to a diode-resistor equivalent circuit model for 200 K. (A colour version of this figure can be viewed online.)

In the ambipolar regime at higher drain voltages (e.g., $V_D = -2$ V), both electrons and holes were injected and accumulated simultaneously, creating an induced p-n junction that was mobile and could be moved through the transistor channel by changing the applied voltages. The corresponding ionic and electronic charge carrier profiles could be fixed, however, by cooling the sample below the melting point of the IL (~240 K) [51] while constant voltages (*e.g.*, $V_D = -2$ V and $V_G = -0.3$ V) were maintained. Thus, a static p-n junction was created that was stable and immobile as long as the IL remained frozen. Importantly, no gate voltage was applied after freezing the ionic liquid.

The frozen p-n junction of the recombination zone led to diode-like I–V characteristics when the drain bias was swept from negative to positive voltages (see Fig. 1c). The corresponding diode rectification, however, was relatively low (~$10^2$) and the current saturation at higher bias ($\geq 2$ V) indicated ohmic effects due to series resistance within the device. A diode-resistor equivalent circuit model [52,53] can be used to fit the I–V curve (see Supplementary Material for details). The resulting fit is shown as a dashed line in Fig. 1c. The deviation from the experimental data in the reverse bias regime probably arises from the inhomogeneity of the SWCNT network, which leads to spatially non-uniform p-n junctions in the relatively wide ($W = 10$ mm) and interdigitated transistor channel. The ideality factor obtained from the fit was 9.3 (*i.e.*, $>2$). Consequently, this p-n junction had to be considered as a non-ideal diode.

### 3.2. Electroluminescence imaging and spectroscopy

The recombination zone of the IL-LEFETs can be positioned and frozen at different distances from the source electrode as evident from the near-infrared electroluminescence images of the channel region in Fig. 2a and b. Note that due to the interdigitated electrode layout, the field of view represents only a small portion of the total channel and the observed width of the emission zone (~1.5 μm) is limited by the resolution of the optical setup. Spectra of the emitted EL recorded at 200 K show clear exciton (E$_{11}$) and trion emission peaks, with the trion peak being red-shifted from the exciton peak by ~165 meV as observed before for doped (6,5) SWCNTs [17]. The ratio of trion to exciton emission varies with applied forward bias and current (see below) but mostly depends on the position of the emission zone (centered or off-center) and thus the distribution of the charge carriers (balanced or unbalanced) around it. In the unbalanced case (Fig. 2a), the ratio of trion to exciton emission is reproducibly higher (~1.4) than in the balanced (Fig. 2b) case (~1.15).

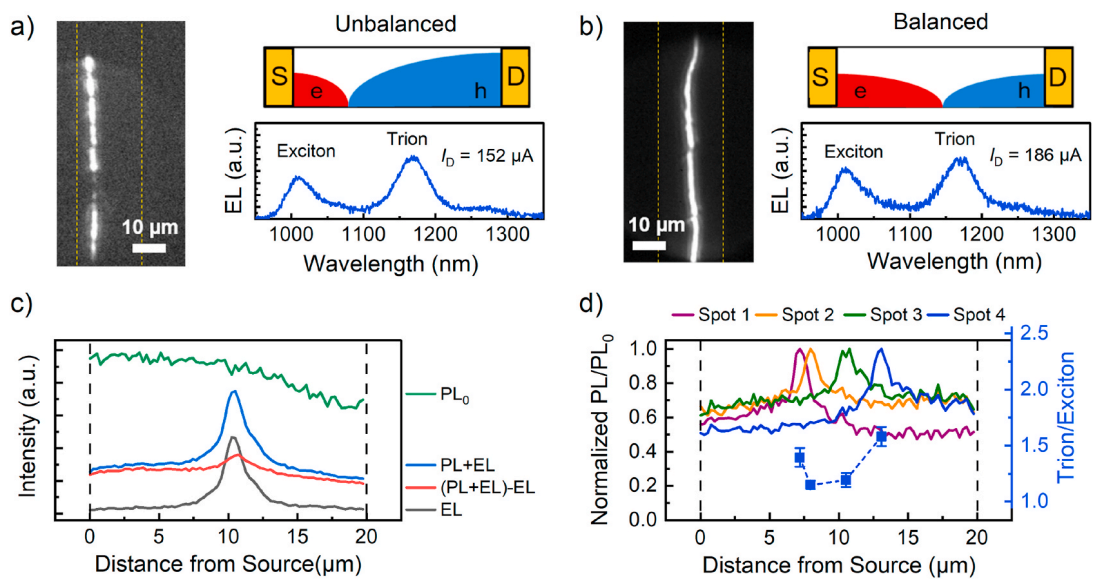

**Fig. 2.** Electroluminescence images (dashed yellow lines indicate the electrode edges) and spectra of IL-LEFETs with static p-n junctions at different positions and schematic illustration of charge carrier density for **a)** off-center (unbalanced, bias applied during cooling, $V_D = -2.1$ V and $V_G = -0.6$ V) and **b)** centered (balanced, bias applied during cooling, $V_D = -2.1$ V and $V_G = -0.3$ V) emission from the channel. **c)** Spatially resolved PL and EL intensity profiles under different conditions: $PL_0$: PL without gating, EL: only electrical excitation, (PL + EL): total emission under optical and electrical excitation, and (PL + EL)-EL: net PL of the p-n junction. **d)** Normalized relative net PL (*i.e.*, $PL/PL_0$) for different positions of the frozen p-n junction and corresponding trion to exciton intensity ratios. All data were recorded at $T = 200$ K. (A colour version of this figure can be viewed online.)

Previous studies had established that the ratio of trion to exciton EL increases with charge carrier density [17], which is non-negligible in the recombination zone of IL-LEFETs [35]. This observation is in contrast to conventional, electrostatically gated LEFETs based on SWCNTs, which barely show any trion EL [37,43]. The dependence of the trion emission on the position of the static p-n junction indicates the impact of excess charge carriers in or close to the recombination zone. This effect can be visualized by recording PL and EL profiles of the transistor channel under different conditions as previously demonstrated for polymer-based LEFETs [54,55] and lateral light-emitting electrochemical cells [56]. The spatially resolved PL from the channel without any applied voltage (and without background) was recorded as PL$_0$, followed by the pure EL signal and then a combined PL and EL signal (PL + EL) with simultaneous optical and electrical excitation (see Fig. 2c). The net PL profile of the channel with a static p-n junction can be defined as (PL + EL) - EL. It shows that the PL is significantly quenched throughout the channel and even to some degree within the p-n junction. This PL quenching is due to the presence of charge carriers that induce Auger-type recombination in addition to trion formation [14,57].

We can estimate the local PL quenching by calculating the ratio of net PL/PL$_0$:

$$\frac{PL}{PL_0} = \frac{(PL+EL) - EL}{PL_0 - BG}, \tag{1}$$

where BG is the background signal. Fig. 2d shows the normalized relative net PL of the channel for different recombination zone positions. For the off-center recombination zones (spots 1 and 4), the normalized PL/PL$_0$ values in the hole (p-type) and electron (n-type) regions are different, corresponding to the different carrier concentrations in these areas. In contrast to that, the hole and electron densities are similar (balanced) for the centered emission zones (spots 2 and 3). These differences in charge excess can be correlated with the ratio of trion to exciton emission in the EL spectra (higher for unbalanced conditions). They suggest that even moderate variations of the charge distribution around a frozen p-n junction directly affect the emission properties. Similar observations were made by Gaulke et al. for short-channel transistors at 4 K with individual or bundles of nanotubes spanning a short channel [28].

### 3.3. Impact of external bias, current density and temperature

To further investigate the role of charge carrier balance for trion emission, current-dependent EL measurements were performed under frozen p-n junction conditions. Note that for larger currents through the frozen p-n junction a higher forward bias must be applied. EL spectra of IL-LEFETs for different drain currents in unbalanced and balanced conditions at 200 K are shown in Fig. 3a and b. The respective integrated intensities of exciton and trion emission were extracted from the EL spectra using a peak fitting routine. The experimental spectra could be reproduced well with two Lorentzian peak functions. Fig. 3c shows the integrated EL intensities of exciton and trion emission for unbalanced biasing conditions during cooling as a function of drain current. Both exciton and trion emission increase with increasing drain current and could be fitted to the equation $EL = A \cdot I_D^m$, where $EL$ is the integrated exciton or trion emission intensity, $A$ is a constant, $I_D$ is the drain current, and $m$ is a power factor. The fitting results show a weak superlinear relation for the trion emission with $m \sim 1.2$ and a linear behavior for the exciton emission with $m \sim 1.0$. A similar trend is observed for balanced conditions, as shown in Fig. 3d, with slightly different power factors of $m \sim 1.0$ for trions and $m \sim 0.8$ for excitons. For unbalanced conditions the trion emission increased relative to the exciton emission with increasing current, while for balanced conditions, the ratio remained nearly constant. Similar trends were also observed at other temperatures (40–200 K) as shown in Fig. 3e. No other significant spectral changes (such as additional peaks or peak shifts) occurred (see Fig. S2, Supplementary Material).

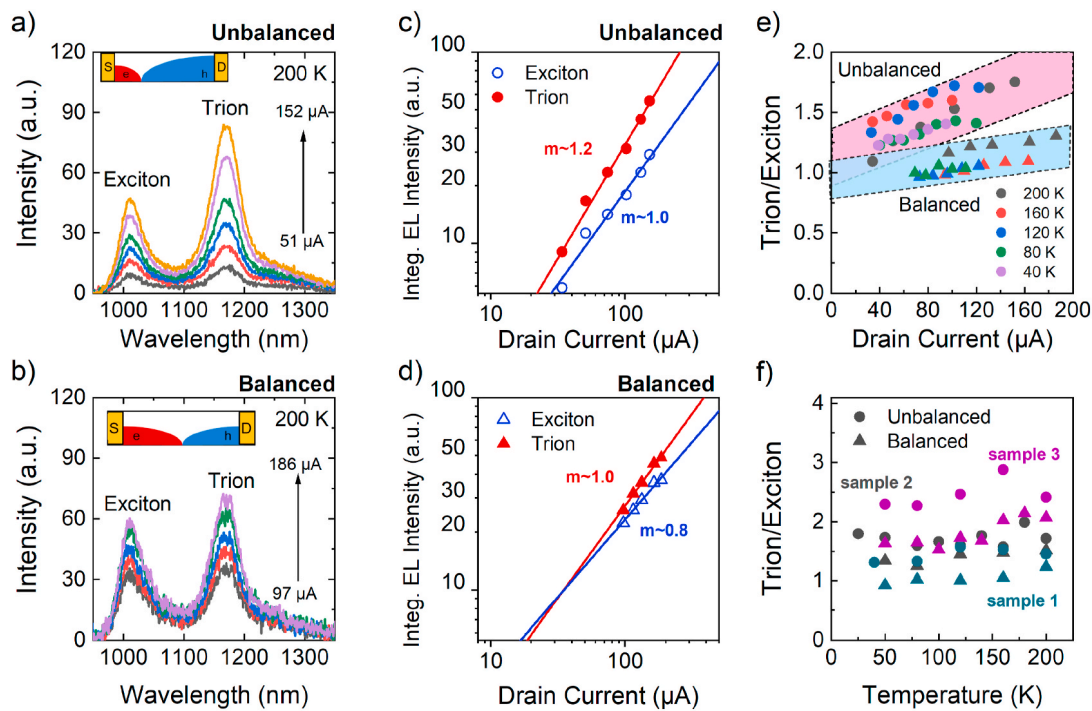

**Fig. 3.** EL spectra from IL-LEFETs under unbalanced (**a**) and balanced (**b**) conditions for different forward currents. Integrated trion and exciton EL intensities depending on drain current for unbalanced (**c**) and balanced (**d**) conditions including the power factor fit (~$I_D^m$). **e**) Trion to exciton EL ratios versus drain current for unbalanced (pink shaded area) and balanced conditions (blue shaded area) at different temperatures. **f**) Temperature-dependent trion to exciton EL ratios for three different samples at $I_D \approx 100$ μA. Drain voltages of $V_D = -2.0$ V, $-2.5$ V, and $-3.0$ V were applied for samples 1, 2, and 3 respectively. Gate voltages $V_G = 0$ V and $-0.3$ V were set for balanced and unbalanced conditions in each sample. (A colour version of this figure can be viewed online.)

The relation between trion and exciton intensity can also be expressed as a power law $EL_{\text{Trion}} = B \cdot (EL_{\text{Exciton}})^{\alpha}$ where the exponent indicates the quasiparticle character with $\alpha \approx 1$ for charged excitons (trions) and $\alpha \approx 2$ for biexcitons [58]. The obtained $\alpha$ values for EL from (6,5) SWCNTs range between 1.1 and 1.3 as shown in Fig. S3 (Supplementary Material), confirming the trion character of the red-shifted EL peak. Fig. 3f shows the trion to exciton ratio for different samples at $I_D \approx 100$ μA depending on temperature. There is no significant temperature dependence of the peak intensity ratio neither for balanced nor for unbalanced conditions. Likewise, no temperature dependence of the trion to exciton ratio was observed for the PL spectra of these networks at different gate voltages (see Fig. S4 and S5, Supplementary Material) and without any induced p-n junction. This absence of any significant impact of temperature suggests that there is no thermal activation mechanism (e.g., via dark excitons) [59] for trion formation and emission.

### 3.4. Model for bias dependent carrier profile

As shown in Fig. 3e the trion to exciton ratio increases more with increasing currents (i.e., external bias) for unbalanced conditions than for balanced conditions, thus indicating more charge carriers in the recombination zone [17]. In conventional gradual channel approximation models for ambipolar transistors, the carrier density in the recombination zone is either zero or balanced [60–62]. These models assume an infinite or very large recombination rate of electrons and holes. All injected holes and electrons recombine within a narrow zone and thus the drain current equals the hole and electron current ($I_D = I_h = I_e$). In reality, the recombination rate is finite [63] and not all of the holes and electrons recombine immediately, leading to an excess of carriers in the recombination zone. These free carriers either quench bright excitons or recombine with excitons to form trions. For IL-LEFETs with mobile recombination zones at room temperature the presence of excess carriers was confirmed by high-resolution in-situ Raman mapping [35], which was not possible for the frozen p-n junctions investigated here.

We now focus on the potential and carrier density profile for balanced and unbalanced conditions and its effect on the ratio of trion to exciton emission. In IL-LEFETs, the induced carrier density is usually on the order of $10^{14}$ cm$^{-2}$ and generates a large built-in electric field across the recombination zone driving carriers into the oppositely charged accumulation layer [17]. By increasing the applied external bias (drain voltage), the electric field increases further and likewise the probability of trion formation. Fig. 4a shows the calculated initial carrier density profile for different ratios of $V_G$ to $V_D$ and thus positions of the recombination zone assuming zero threshold voltage, equal hole and electron mobilities, and no contact resistance (for details see Supplementary Material). The concentration of excess electrons and holes in the proximity of the recombination zone is always the same and carrier balance is preserved, as indicated by the dashed line in the red-highlighted zones.

If we, however, consider a static p-n junction formed by freezing the ionic liquid under certain bias conditions, additional assumptions are required. First, the carrier density profile becomes fixed and the recombination zone cannot move with bias changes. A new condition $x' = x_0$ is added, where $x_0$ and $x'$ are the distances of the recombination zone from the source electrode before and after an external bias $V_{\text{ext}}$ is applied, respectively. Second, the potential difference under external bias generates an external electric field $\vec{E}_{\text{ext}}$ across the p-n junction (here assumed to be constant). Thus, the carrier density under external electric field $N'(x)$ can be written as

$$N'(x) = N_0(x) + C\left(\frac{V_{\text{ext}}}{L}x - \frac{V_G}{V_D}V_{\text{ext}}\right), \qquad (2)$$

where $V_D$ and $V_G$ are the initial drain and gate voltages, $C$ is the capacitance, $L$ is the channel length, and $N_0(x)$ is the initial carrier density (for derivation see Supplementary Material). The capacitance in equation (2) is the electric double layer capacitance before the ions are immobilized. Note that $V_G$ is the externally applied gate voltage assuming no additional voltage drops due to the electrical double layer capacitances at the source, drain, and gate electrodes. This value is related but not equal to the potential in the ionic liquid. A more precise potential could be determined by including a reference electrode or by equivalent circuits if the three capacitances were known.

Fig. 4b shows the carrier density profiles for unbalanced conditions in the proximity of the emission zone as obtained from equation (2). The solid black and dashed curves indicate the carrier density with and without applied external field within the static p-n junction. The external bias clearly tilts the carrier density profile of the junction and creates an unbalanced distribution of holes and electrons around the recombination zone. The resulting excess of carriers in the recombination zone favors trion formation. Note that in this model, the excess carriers in forward bias are electrons and thus emission from negative trions would be expected. Due to the similarity of the binding energy of negative and positive trions (only few meV difference) [13,17] and broad emission peaks it is not possible to unambiguously assign the experimentally observed trion EL.

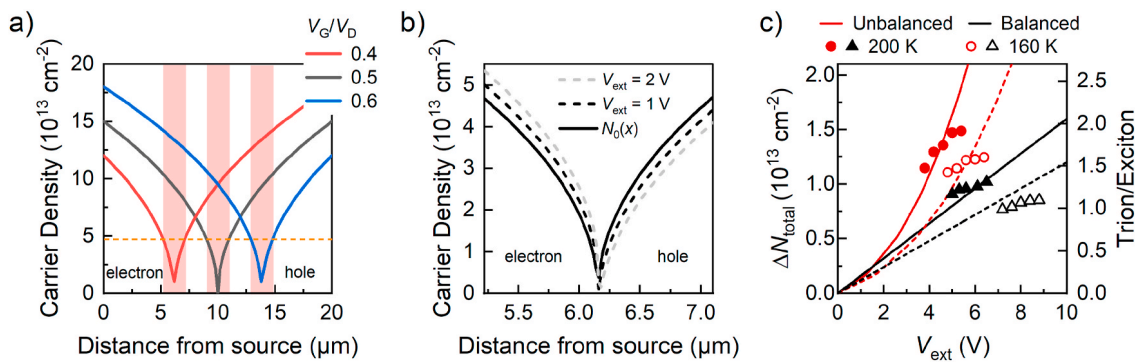

**Fig. 4. a)** Carrier density profiles calculated from gradual channel approximation under unbalanced (red and blue lines) and balanced (black line) conditions with $V_D = 3$ V. The highlighted areas indicate the apparent emission zones (width $\sim 2$ μm). **b)** Carrier density profile in the proximity of the emission zone under unbalanced condition calculated from equation (2), initial electron-hole profile ($N_0(x)$, solid black line), and with applied external voltage across the channel ($N'(x)$, dashed lines). **c)** Trion/exciton intensity ratios (closed symbols: 200 K; open symbols: 160 K) obtained from EL experiment depending on applied external bias. Calculated total excess carrier densities $\Delta N_{\text{total}}$ (integrated across the recombination zone: ±1 μm from the center of the recombination zone) versus applied external voltage ($V_{\text{ext}}$) for balanced (black lines) and unbalanced (red lines) conditions. The solid and dashed lines indicate $\Delta N_{\text{total}}$ for p-n junctions in IL-LEFETs with capacitances of 3.2 μF cm$^{-2}$ and 2.5 μF cm$^{-2}$, respectively. Note, they are not fits to the experimental data points. (A colour version of this figure can be viewed online.)

To examine the effect of the initial carrier balance conditions on the total carrier density in the recombination zone, we compare the calculated total excess carrier density for balanced and unbalanced conditions in the recombination zone as a function of external bias (see Fig. 4c). The total density of excess carriers is $\Delta N_{total} = N' - N_0$, where $N'$ and $N_0$ are the total carrier densities integrated across the recombination zone, i.e., within $\pm 1$ μm from the center, under external bias and without bias, respectively. The increase of excess carriers in the recombination zone for unbalanced conditions (red lines) is superlinear (almost quadratic), while for balanced conditions a linear relation (black line) is obtained. A similar trend is observed when we compare the results of this simple model to the trion to exciton ratios (symbols in Fig. 4c) at 200 K and 160 K, indicating that the increase of trion emission with drain current (i.e., applied external bias) arises from the local excess of carriers in the recombination zone. Furthermore, the stronger dependence of the trion to exciton ratio for unbalanced versus balanced conditions as shown in Fig. 3e can be explained with this simple model. Note, that a spatial resolution of trion versus exciton emission within the emission zone (width ~1.5 μm) was not possible due to the fundamental limitations of optical microscopy in the near-infrared.

*3.5. Electroluminescence yield*

Finally, the EL efficiency of such IL-LEFETs should depend on Auger quenching by excess charges. Previous studies had estimated an EL efficiency of less than $10^{-5}$ photons per injected electron in IL-LEFETs, which is substantially lower than for electrostatically gated LEFETs that do not show significant trion emission [17,37]. Fig. S6 (Supplementary Material) shows that the EL efficiency is low but nearly constant in the low current regime ($I_D < 200$ μA). It decreases in the high current regime ($I_D > 200$ μA) irrespective of initial conditions (balanced or unbalanced), which indicates the increasing charge excess in the recombination zone with external bias in both cases. Overall, the large excess carrier concentrations in this type of device are highly detrimental for efficient light emission. Interestingly, multilayer organic light-emitting diodes with (6,5) SWCNTs as the emissive layer also featured a static p-n junction and large external bias across this junction. Likewise, these devices showed almost only trion emission and low external quantum efficiencies (<0.01%) [27].

## 4. Conclusions

We have demonstrated the formation of static p-n junctions in ionic liquid-gated (6,5) SWCNT networks. The immobile p-n junctions were created by cooling down the devices below the melting point of the ionic liquid [EMIM][FAP] while continuously applying drain and gate voltages corresponding to the ambipolar regime. The frozen junctions showed diode-like current-voltage rectification and near-infrared electroluminescence with almost equal contributions of exciton and trion emission. The precise ratio varied with the initial conditions of p-n junction formation (balanced or unbalanced) and the applied external bias. However, it did not depend on temperature (30–200 K). Stronger trion emission mainly resulted from unbalanced carrier concentration profiles with excess charge carriers within the recombination zone. An adapted equation for the potential and carrier density profiles in ambipolar transistors with static p-n junctions was used to calculate the number of excess charges in the recombination zone depending on the applied external bias. This simple model can reproduce the larger dependence of the trion to exciton ratio on external bias for unbalanced p-n junctions (i.e., recombination zone frozen close to one of the electrodes) compared to balanced junctions (in the middle of the channel). As the presence of excess carriers and thus quenching is one of the main reasons for the low electroluminescence efficiencies of SWCNT-based light-emitting devices it should be avoided or limited as much as possible. However, the clear spectroscopic signature of trions also makes them sensitive indicators for charge imbalances and excess carriers within optoelectronic devices as shown here.

## CRediT authorship contribution statement

**Abdurrahman Ali El Yumin:** Conceptualization, Methodology, Investigation, Data curation, Formal analysis, Writing – original draft, Writing – review & editing. **Nicolas F. Zorn:** Methodology, Investigation, Formal analysis, Writing – original draft, Writing – review &. **Felix J. Berger:** Investigation, Writing – review & editing. **Daniel Heimfarth:** Investigation, Writing – review & editing. **Jana Zaumseil:** Conceptualization, Methodology, Funding acquisition, Resources, Writing – review & editing.

## Declaration of competing interest

The authors declare that they have no known competing financial interests or personal relationships that could have appeared to influence the work reported in this paper.

## Acknowledgments

This project has received funding from the European Research Council (ERC) under the European Union's Horizon 2020 research and innovation programme (Grant agreement no. 817494 "TRIFECTs").

## Appendix A. Supplementary data

Supplementary data to this article can be found online at https://doi.org/10.1016/j.carbon.2022.11.025.

# Supplementary Material

# Trion emission from frozen p-n junctions in networks of electrolyte-gated (6,5) single-walled carbon nanotubes

Abdurrahman Ali El Yumin, Nicolas F. Zorn, Felix J. Berger, Daniel Heimfarth and Jana Zaumseil*

*Institute for Physical Chemistry, Universität Heidelberg, 69120 Heidelberg, Germany*

\* Corresponding Author:

Phone: +49-6221-54 5065, E-mail: zaumseil@uni-heidelberg.de

# Contents

## (6,5) SWCNT Dispersion and Film Characterization

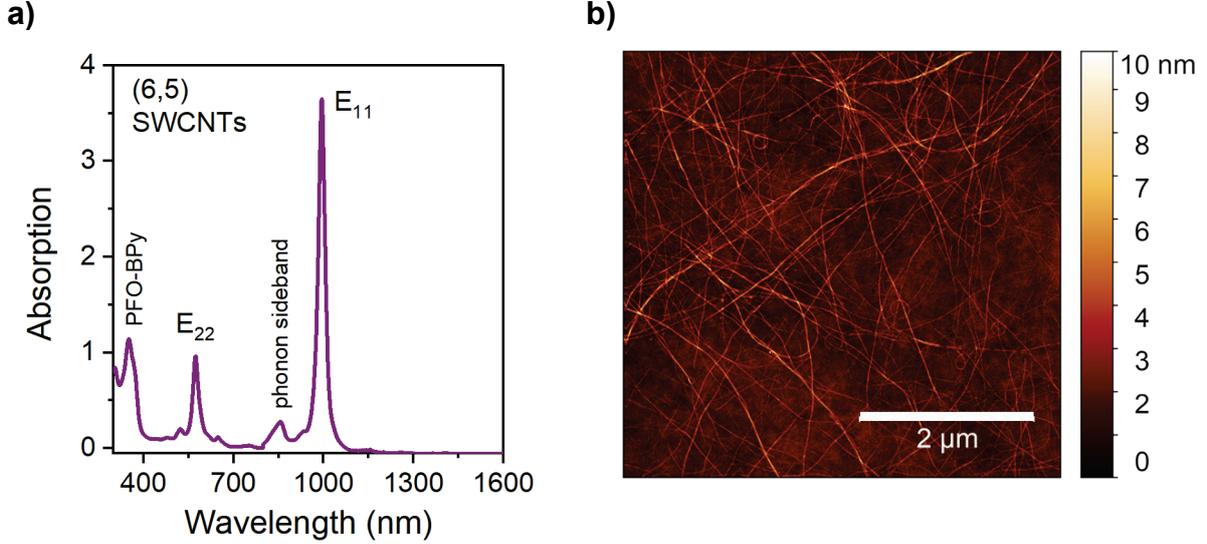

**Figure S1.** a) Typical absorption spectrum of PFO-BPy-sorted (6,5) single-walled carbon nanotubes in toluene. b) Representative atomic force microscopy (AFM) image (ScanAsyst mode) of a spin-coated (6,5) SWCNT network as used for IL-LEFET fabrication.

## Fitting Diode *I-V* Curves of Frozen p-n Junctions

A diode-resistor equivalent circuit model comprising a diode, a series resistance element and a parallel shunt resistance element can be used to fit the *I-V* characteristics of the frozen p-n-junction (see Figure 1c). The diode equation can be written as [1, 2]

$$I_D = I_0 \left\{ \exp\left[\frac{q(V_D - I_D R_S)}{nkT}\right] - 1 \right\} + \frac{V_D - I_D R_S}{R_{sh}}, \qquad (1)$$

where $I_D$ and $V_D$ are drain current and voltage, $I_0$ is the diode reverse current, $n$ is the diode ideality factor, $q$ is an electron charge, $k$ is the Boltzmann constant, $T$ is the temperature, $R_s$ is an effective series resistance, and $R_{sh}$ is a parallel shunt resistance. The exact analytical solution of equation (1) has been shown by Ortis-Conde *et al.* [2]. From the fitting procedure, we obtain $I_0 \approx 3.15$ pA, $R_s \approx 30$ kΩ, $R_{sh} \approx 120$ MΩ, and $n \approx 9.3$ at $T = 200$ K. Equation (1) fits well with the data in the forward bias regime ($I_D > 0$) although we observe a clear discrepancy in the reverse bias regime ($I_D < 0$) where the reverse bias current is higher than predicted. Furthermore, the obtained ideality factor $n \approx 9.3$ is much higher than that of ideal diodes with ideality factors from 1 to 2. Diodes with ideality factors $n > 2$ are commonly regarded as a non-ideal. This might arise from the inhomogeneity of the SWCNT network, which leads to spatially non-uniform p-n junction generation in a wide transistor channel ($W = 10$ mm) and is not included in this model. Several studies on single SWCNTs with split-gate p-n junctions showed ideality factors of $n \approx 1$ [3, 4]. However, in the case of a nanotube network, the inter-tube junctions can affect the diode performance since charge transport in the SWCNT network is mainly limited by carrier hopping between individual nanotubes. The presence of inter-nanotube junctions in the recombination zone may reduce the diode efficiency.

Finally, $R_s \approx 30$ k$\Omega$ represents the effective series resistance element outside the recombination zone including the bulk-resistance resulting from the inter-nanotube (p- and n-type regions) junctions and the contact resistance at the electrodes. This series resistance becomes more pronounced at lower temperatures due to lower carrier mobilities within the network and increased contact resistance [5] leading to a decrease of diode current.

# Current-Dependent Electroluminescence Spectra

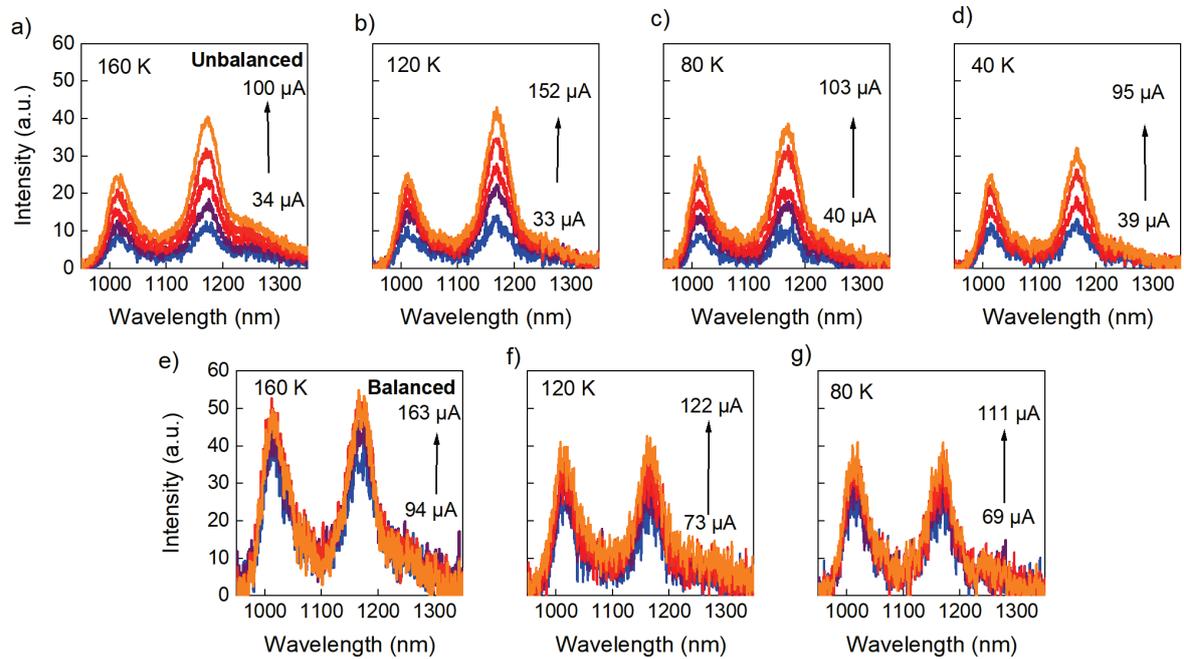

**Figure S2.** Current-dependent EL spectra from IL-LEFETs with frozen p-n-junctions at different temperatures. Upper row (a-d): unbalanced condition, lower row (e-g): balanced condition.

# Trion to Exciton Intensity Relation

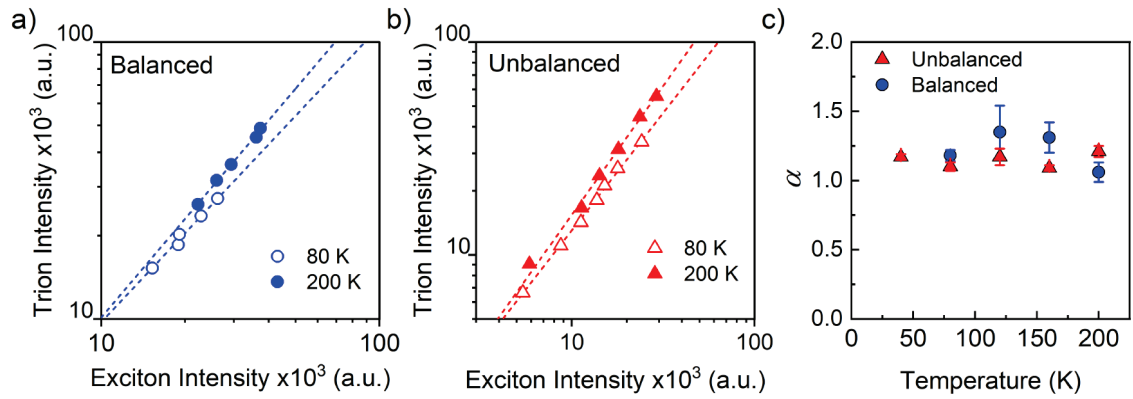

**Figure S3.** Correlation between trion and exciton intensity at different drain currents expressed as a power law $EL_{\text{Trion}} = B \cdot (EL_{\text{Exciton}})^\alpha$ for balanced (a) and unbalanced (b) conditions. c) Power factor $\alpha$ at different temperatures. The error bars represent the standard deviation of the fitting results. No explicit temperature dependence of $\alpha$ is observed.

## Temperature-Dependent Photoluminescence

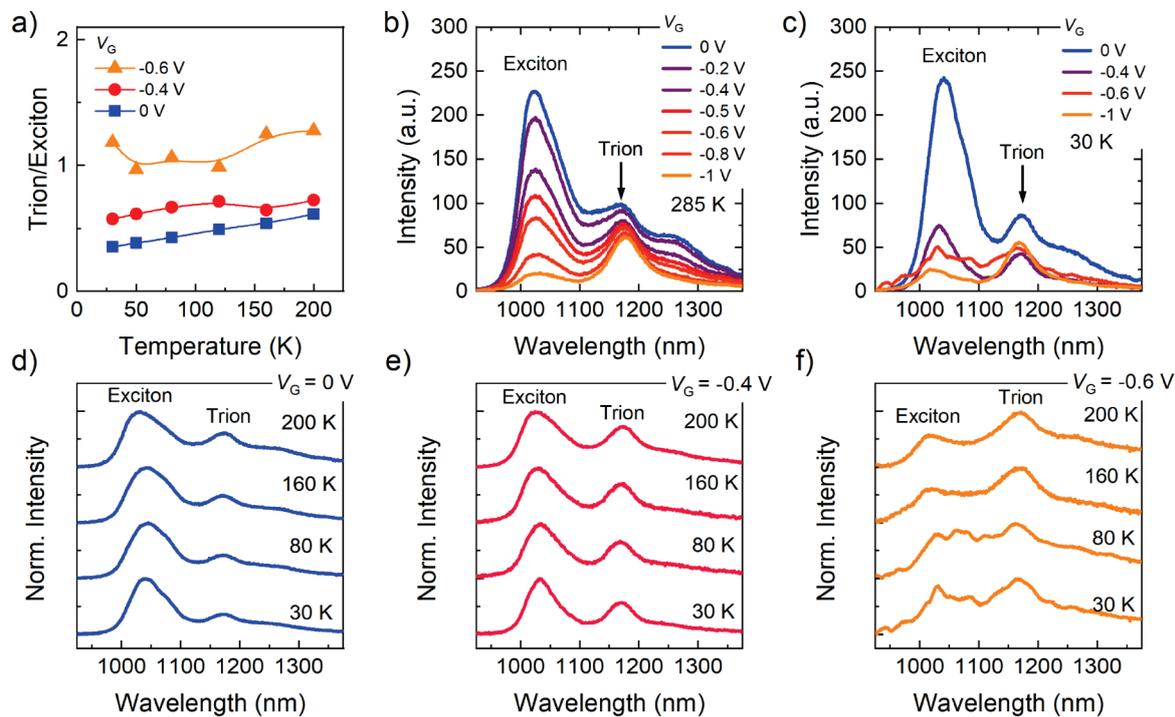

**Figure S4.** a) Temperature-dependent trion to exciton photoluminescence (PL) intensity ratios at different applied gate voltages ($V_G$) for hole accumulation (no drain bias, no p-n junction). b) $V_G$-dependent PL at $T = 285$ K. c) $V_G$-dependent PL at $T = 30$ K. To obtain low temperature $V_G$-dependent spectra, $V_G$ was applied at $T = 285$ K and while cooling down to 30 K before acquiring PL spectra. (d-e) Temperature-dependent spectra for different applied gate voltages.

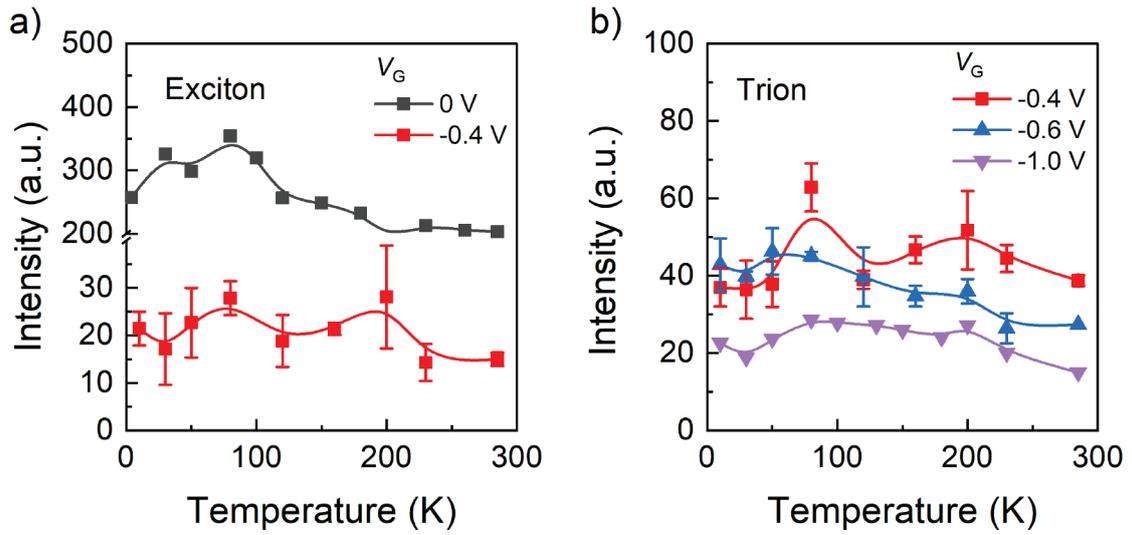

**Figure S5**. a) Temperature-dependent integrated intensities for excitonic PL at $V_G = 0$ V (gray squares) and $V_G = -0.4$ V (red squares). Solid lines are guides for the eye. b) Temperature-dependent integrated intensity of trion PL at different applied gate voltages. Note, the excitonic PL at $V_G = -1$ V was completely quenched and could not be resolved. No significant temperature dependence was observed in either case.

**Estimation of Carrier Density Profiles under an External Field**

The carrier density $N(x)$ along the transistor channel in the ambipolar regime can be expressed as [6, 7]

$$N(x) = C(V_G - V(x)) \qquad (2)$$

$$V(x) = V_G \begin{cases} \pm \sqrt{V_G^2 - \frac{x}{x_0} \cdot V_G^2} & \text{for } x \leq x_o \\ \mp \sqrt{(V_D - V_G)^2 - \frac{(L-x)}{(L-x_0)} \cdot (V_D - V_G)^2} & \text{for } x \geq x_o \end{cases} \qquad (3)$$

where $L$ is channel length, $V_G$ and $V_D$ are gate and drain voltages, $x_0$ is the distance (position) of the recombination zone from the source electrode, and $C$ is electrical double layer (EDL) capacitance. Note that $V_G$ is the externally applied gate voltage assuming no additional voltage drops due to the electrical double layer capacitances at the source, drain, and gate electrodes. This value is related but not equal to the potential in the ionic liquid. A more precise potential could be determined by including a reference electrode or by equivalent circuits if the three capacitances were known.

Furthermore, the position of recombination zone can be written as

$$x_0 = L \frac{V_G^2}{V_G^2 + \frac{\mu_h}{\mu_e}(V_G - V_D)^2}. \qquad (4)$$

For static conditions, the recombination zone is fixed thus preventing any recombination zone movement when the lateral bias changes. Therefore, we introduce a new condition

$$x' = x_0 \qquad (5)$$

where $x_0$ and $x'$ are the distance of the recombination zone from the source electrode before and after the external bias $V_{ext}$ is applied. Once the ionic liquid is frozen, the ions remain in place even when the gate bias is removed, hence the potential and carrier density profile are retained. Combining equation (4) and (5), we obtain:

$$V_G' = \frac{V_G}{V_D} V_D'. \qquad (6)$$

if we replace $V_D'$ with $V_D + V_{ext}$, equation (6) becomes

$$V_G' = V_G + \frac{V_G}{V_d}V_{ext}. \tag{7}$$

The total electric field close to the recombination zone can thus be described as:

$$\vec{E}_{tot}(x) = \frac{dV_0(x)}{dx} + \vec{E}_{ext}, \tag{8}$$

where $V_0(x)$ is the initial potential profile. Note that $\vec{E}_{ext}$ can also be expressed as $V_{ext}/L$, where $V_{ext}$ is external bias voltage. Using a simple integration, we obtain the potential and carrier densities profile under external field for a static p-n junction:

$$V(x) = V_0(x) + \frac{V_{ext}}{L}x, \tag{9}$$

$$N'(x) = C(V(x) - V_G'). \tag{10}$$

By substituting $V_G'$ in equation (10) with equation (7), the carrier density under external field becomes:

$$N'(x) = N_0(x) + C\left(\frac{V_{ext}}{L}x - \frac{V_G}{V_D}V_{ext}\right) \tag{11}$$

# Estimated Electroluminescence Efficiency

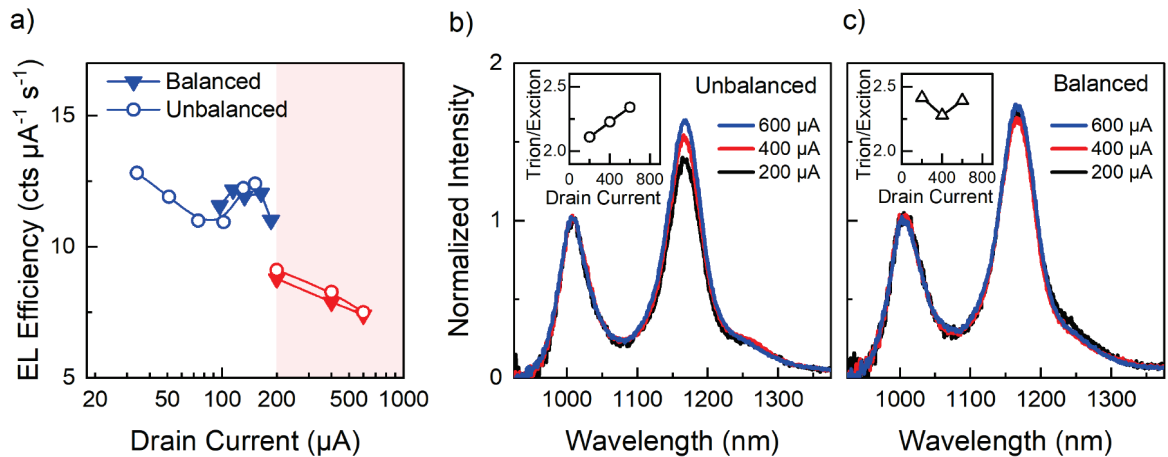

**Figure S6.** a) Dependence of overall EL efficiency on drain current at 200 K. The red-highlighted area represents the high current regime in which the p-n junction was formed by applying $V_D$ = -2.5 V and gate voltages were set as $V_G$ = 0 V (balanced) and -0.42 V (unbalanced) at room temperature before cooling down. For the low current regime, $V_D$ = -2.1 V with $V_G$ = 0 V (balanced) and -0.3 V (unbalanced) were set to form the p-n junction. The corresponding EL spectra (normalized to exciton emission) in the higher current regime are shown in (b) and (c) for unbalanced and balanced conditions, respectively. Inset: the extracted trion to exciton ratio at different current densities.